\documentclass[12pt]{article}

 \UseRawInputEncoding 

\usepackage{color}
\usepackage[backref=page]{hyperref}
\hypersetup{colorlinks=true,linkcolor=blue}

\newtheorem{theorem}{Theorem}[section]
\newtheorem{lemma}[theorem]{Lemma}
\newtheorem{corollary}[theorem]{Corollary}
\newtheorem{definition}[theorem]{Definition}

\newcommand\prob{{\mbox Prob}}
\newcommand\php{{\mbox{PHP}}}
\newcommand\pphp{{\neg{{\cal P}{\cal H}{\cal P}}}}
\newcommand\fa{{\cal F}}
\newcommand\ea{{\cal E}}
\newcommand\ff{{\mathbf F}}
\newcommand\fp{{{\ff}_p}}
\newcommand\rx{{\fp[\overline x]}}

\newcommand\re{{\cal R}({\cal E})}
\newcommand\rd{{\fp^{\le d}[\overline x]}}

\newcommand\phpd{{\fp^{\le d}[Var({\pphp}_n)]}}
\newcommand\one{{\Omega(n,d)}}
\newcommand\fn{{\cal F}_n}

\newcommand\fl{F_{\ell}(\mbox{MOD}_p)}
\newcommand\des{{\sf Des}}
\newcommand\spa{{\sf Span}}
\newcommand\res{{\sf Res}}

\newcommand\errho{{\mbox{Error}(\rho)}}
\newcommand\labp{{\mbox{lab}(P)}}

\newcommand\qed{\begin{flushright} {\bf q.e.d.} \end{flushright} }
\newcommand\prf{\noindent {\bf Proof :}}

\begin{document}

\title{Pseudo-solutions of polynomial systems\\
	 and the lower bound problem for\\
$\mbox{AC}^0[p]$-Frege systems}

\author{Jan Kraj\'{\i}\v{c}ek}

\date{Faculty of Mathematics and Physics\\
Charles University\thanks{
Sokolovsk\' a 83, Prague, 186 75,
The Czech Republic, {\tt jan.krajicek@protonmail.com}}}

\maketitle

\begin{abstract}
	
The problem to establish a lower bound for $\mbox{AC}^0[p]$-Frege refutations of 
a system of polynomial equations over $\fp$ was in \cite{Kra-finitary} reduced 
to the existence of a pseudo-solution (a notion defined there) for the system.
	
Here we reduce this further, for the system expressing the negation of the PHP, 
to a property of search trees querying values of linear maps
on the vector space of low degree polynomials over $\fp$.

\end{abstract}

\section*{Introduction}

$\mbox{AC}^0[p]$-Frege systems are propositional proof calculi that operate with
constant depth formulas in the DeMorgan language augmented by connectives counting
modulo prime $p$. The problem to establish lengths-of-proofs lower bounds for these
calculi runs through proof complexity over thirty years and 
numerous developments in the field relate directly or indirectly to it. 

The problem was inspired by a success in proving super-polynomial \cite{Ajt88} 
and exponential \cite{Kra-lower,KPW,PBI} lower bounds for $\mbox{AC}^0$-Frege systems. 
Proofs of these lower bounds were in part based on modifications of the random 
restriction method (and switching lemmas) from circuit complexity. It was thus
natural to think that one ought to be able to modify also the method of 
lower bounds for $\mbox{AC}^0[p]$-formulas of \cite{Raz87,Smo}
to proof complexity too. 
The first step was achieved in \cite{BIKPRS} where a proof-theoretic counterpart 
of approximating $\mbox{AC}^0[p]$-formulas by low degree polynomials was defined.
It consisted of introducing a new algebraic proof system 
(Extended Nullstellensatz proof system ENS) that operates with low degree polynomials 
and showing that it is equivalent (in the sense of simulations) to $\mbox{AC}^0[p]$-Frege
systems.
The second step then ought to be proving a lower bound for ENS; however, this 
step has proved to be rather elusive. In \cite[Sec.5]{Kra-finitary} we reduced
the lower bound to ENS-refutations of a polynomial system to the task to construct its
suitable pseudo-solution (a notion invented there). 

A candidate pseudo-solution for
system $\pphp_n$ expressing the negation of the PHP was proposed in \cite{Kra-finitary}
and whether or not it has the required properties was left as an open problem,
cf. \cite[Problem 4.4]{Kra-finitary}.
In this paper we reduce this problem further to the task to establish a mild 
(sub-exponentially small) lower
bound for the error particular search trees solving a problem about PHP-type 
restrictions must make.

To make the paper reasonably self-contained we recall all definitions and statements
(but not their proofs) from \cite{BIKPRS} and \cite{Kra-finitary} that are needed
to understand the argument:
the definitions of $\mbox{AC}^0[p]$-Frege systems and the $\php_n$ formulas and 
the statement of the lower bound are given in Section \ref{formulation}, 
the ENS proof system and the reduction of the 
$\mbox{AC}^0[p]$-Frege systems lower bound
problem to a lower bound problem for ENS
is recalled in Section \ref{ens}, and 
pseudo-solutions and a further reduction to the problem to their existence 
from \cite{Kra-finitary} are recalled in Section \ref{reduct}.
Suitable pseudo-solutions for $\php_n$ are defined in Section \ref{php} and 
their properties are investigated in Section \ref{key1}. In Section \ref{key2}
we formulate the main result: the reduction of
a lower bound for $\mbox{AC}^0[p]$-Frege refutations of $\pphp_n$ to
a specific task to lower bound a certain combinatorially defined probability
involving search tress and linear maps. The paper is concluded
by a remark in Section \ref{remarks}.

\medskip

Because of the long history of the problem in proof complexity there are many
papers that touch it in one way or other. 
Here we give only references that are directly relevant to 
our paper and we leave it to the reader to consult \cite{prf} for more background.

\medskip

\noindent
{\bf Conventions:} For $n \geq 1$ we let $[n] := \{1, \dots, n\}$ and we fix 
an arbitrary prime $p$ for the rest of the paper.

\section{Preliminaries} \label{formulation}

$\mbox{AC}^0[p]$ proof systems
are best defined as sequent calculus with unbounded fan-in conjunctions, disjunctions and 
modular counting gates. However, the papers \cite{BIKPRS,Kra-finitary} we build on 
used Frege systems and we shall thus stick to that choice. 

For $F$ a Frege system in the DeMorgan language $0, 1, \neg, \vee,\wedge$
denote by
$F(\mbox{MOD}_p)$ the proof system obtained from $F$ by adding
unbounded arity connectives $\mbox{MOD}_{p,i}$ for $i = 0, \dots, p-1$
to the language and some new axioms that reflect the interpretation that
$\mbox{MOD}_{p,i}(y_1, \dots, y_k)$ is true iff $\sum_j y_j \equiv i\ (\mbox{mod } p)$. 
The following set of $\mbox{MOD}_p$-axioms was adopted in \cite[Sec.12.6]{kniha}
and \cite{BIKPRS}:
\begin{itemize}
	\item $\mbox{MOD}_{p,0}(\emptyset)$
	
	\item $
	\neg \mbox{MOD}_{p,i}(\emptyset)\ ,\ \mbox{for}\ i=1,\dots, p-1
	$
	
	\item $
	\mbox{MOD}_{p,i}(\Gamma,\phi) \ \equiv\ [(\mbox{MOD}_{p,i}(\Gamma) \wedge
	\neg \phi) \vee (\mbox{MOD}_{p,i-1}(\Gamma) \wedge \phi)]
	$
	
	for $i = 0, \dots, p-1$, where $0-1$ means $p-1$ 
	and where $\Gamma$ stands for any sequence (possibly empty)
	of formulas.
\end{itemize}
The {\bf depth} of a formula is defined inductively:
the depth of a constant or of an atom is $0$, the use of the negation or of any
of $\mbox{MOD}_{p,i}$ increases the depth
by $1$, and a formula formed from formulas $A_i$ none of which starts with $\vee$ (resp. with $\wedge$)
by a repeated use of $\vee$ (resp. of $\wedge$) has the depth $1$ plus the maximum depth
of formulas $A_i$. 
Given $\ell \geq 0$, $\fl$ is the subsystem of $F(\mbox{MOD}_p)$ whose proofs
may contain only formulas of depth at most $\ell$ 

\medskip

For $n \geq 1$ the pigeonhole principle tautology $\php_n$ introduced in \cite{CooRec} 
is:
$$
\bigvee_{i} \bigwedge_j \neg p_{ij}\ \vee\  
\bigvee_{i_1\neq i_2, j} (p_{i_1 j} \wedge p_{i_2 j})\ \vee\  
\bigwedge_{i, j_1 \neq j_2} (p_{i j_1} \wedge p_{i j_2})
$$
with $i, i_1, i_2$ ranging over $[n+1]$ and $j, j_1, j_2$ over $[n]$.
The formula expresses that the condition 
$p_{i j} = 1$ cannot define the graph of an injective function
from $[n+1]$ into $[n]$ (the pigeonhole principle).

In order to take an advantage of algebraic proof systems replace (following \cite{BIKPP}) 
the formula $\neg \php_n$ by an unsolvable system of polynomial equations
over $\fp$ in variables $\{x_{i j}\ |\ i \in [n+1], j \in [n]\}$:
\begin{itemize}
	
	\item $x_{i_1 j} \cdot x_{i_2 j} = 0$,  
	for each $i_1 \neq i_2 \in [n+1]$ and $j \in [n]$.
	
	\item $x_{i j_1} \cdot x_{i j_2} = 0$,  
	for each $i \in [n+1]$ and $j_1 \neq j_2 \in [n]$.
	
	\item $1 - \sum_{j \in [n]} x_{i j} = 0$, for each $i \in [n+1]$.
	
	\item $x^2_{i j} - x_{i j}$, for all $i \in [n+1], j\in [n]$.
	
\end{itemize}
Denote the left-hand sides of the equations in the first three items 
as $Q_{i_1, i_2; j}$,
$Q_{i; j_1, j_2}$ and $Q_i$, respectively, and let $\pphp_n$ be the 
set of all these polynomials together with all $x_{i j}^2 - x_{i j}$.
Denote also by $Var(\pphp_n)$ the set of 
variables occurring in these polynomials.

Monomials in a polynomial can be defined by conjunctions
and hence equation $f=0$ for a polynomial $f$ over $\ff_p$ is, in particular, also a
depth $2$ formula in the language of $F(\mbox{MOD}_p)$.
Therefore it makes sense to talk about $F(\mbox{MOD}_p)$-refutations of $\pphp_n$.

\section{Reduction to ENS} \label{ens}

Recall from \cite{BIKPP} that  
an {\bf NS-refutation} of a set of polynomials $\cal F$ with variables $Var({\cal F})$ and
containing all polynomials $x^2 - x$ for all $x \in Var({\cal F})$ is a tuple of polynomials
$h_f \in \fp[Var(\fa)]$, for $f \in {\cal F}$, such that
$$
\sum_{f \in {\cal F}} \ h_f \cdot f\ =\ 1
$$
holds in $\fp[Var({\cal F})]$. The {\bf degree of the refutation} is 
$$
\max_{f \in {\cal F}} deg(h_f f)\ .
$$

This refutation system can be further strengthened by extending the initial set
${\cal F}$ suitably.

\begin{definition} [\cite{BIKPRS}] \label{2.1}
	{\ }

	Let $\cal F$ be a system of polynomials.
	\begin{enumerate}
		
		\item Let $\overline g = g_1, \dots, g_m$ be polynomials in any variables over $\ff_p$ and let $h \geq 1$ be a parameter.
		For $i \le m$ define polynomials:
		\begin{equation} \label{e4}
			E_{i,\overline g}\ :=\ g_i \cdot \Pi_{u \le h} (1 - \sum_{j \le m} r_{u j} g_j)
		\end{equation}
		where $r_{u j}$ are new {\bf extension variables} common to all $i\le m$.
		The polynomials $E_{i,\overline g}$ are called the
		{\bf extension polynomials of accuracy $h$} corresponding to $\overline g$.
		
		\item A set $\cal E$ of extension polynomials 
		can be {\bf stratified into $\ell$ levels} iff $\cal E$ can be partitioned
		as ${\cal E}_1 \cup \dots \cup {\cal E}_\ell$ where:
		
		\begin{itemize}
			
			\item If $E_{i, \overline g} \in {\cal E}_t$, some $t$ and $i$, then also
			all companion polynomials $E_{j, \overline g}$ are in ${\cal E}_t$, all $j \le m$.
			
			\item Variables in the polynomials $g_j$ in $E_{i, \overline g}$
			in ${\cal E}_1$ are among $Var({\cal F})$
			and no extension variable from $E_{i, \overline g}$
			occurs among $Var({\cal F})$ or in other extension polynomials in ${\cal E}_1$
			except in the companion polynomials.
			
			\item Variables in the polynomials $g_j$
			in the axioms $E_{i,\overline g}$ in ${\cal E}_{t+1}$, $1 \le t < \ell$, 
			are among the variables occurring in $Var({\cal F} \cup\ \bigcup_{s \le t} {\cal E}_s)$ (including the extension variables from these levels)
			and no extension variable from $E_{i,\overline g}$ does occur among them or in other extension polynomials 
			in ${\cal E}_{t+1}$ except in the companion polynomials.
			
		\end{itemize}
		\item For a set $\cal E$ of extension polynomials put
		\begin{itemize}
			\item ${\cal R}({\cal E})$ to be the
			set of polynomials $r^p - r$, for all extension variables $r$ occurring in $\cal E$.
		\end{itemize}
	\end{enumerate}
\end{definition}
An {\bf ENS-refutation} of $\fa$ consists of a triple $(h, \ea, L)$, where 
$h \geq 1$ is its accuracy, $\ea$ is a set satisfying 
Definition \ref{2.1} and $L$ is an NS-refutation of $\fa \cup \ea \cup \re$. Its degree
is the degree of $L$.

The following theorem is a part of a more general theorem proved in \cite{BIKPRS};
we specialize it to $\pphp_n$.

\begin{theorem} [{\cite[Thm.6.7(1)]{BIKPRS}}] \label{bikprs}
	{\ }
	
	For any constant $\ell \geq 2$, any parameter $h \geq 1$ and any $n \geq 1$: 
	if ${\pphp}_n$ has an 
	$F_\ell(\mbox{MOD}_p)$-refutation with $k$ steps then there is an ENS-refutation 
	of $\pphp_n$ with 
	$S = k^{O(1)}$ number of extension polynomials stratified into $\ell + O(1)$ levels, 
	accuracy $h$ and of degree
	$$
	d \le (2 + \log k)(h+1)^{O(\ell)}\ .
	$$
	The constants implicit in the O-notation depend only on $F$ and $p$.
\end{theorem}

\section{Reduction to pseudo-solutions} \label{reduct}

Let $\cal F$ be a polynomial system over $\fp$ with variables $\overline x$.
For $d \geq 0$ let $\rd$ be the $\fp$-vector space of polynomials from $\rx$ of degree
at most $d$.
An {\bf $\rd$-tree} $T$ is a finite 
$p$-ary tree whose each non-leaf vertex is labeled by a query $g =\ ?$ for some
polynomial $g\in \rd$
and the $p$ outgoing edges are labeled by $g = a$, for all $a \in \fp$. The leaves
of $T$ are labeled by elements of some non-empty set $I$. 
The {\bf height of $T$} is the maximum number of edges on a path from the root to
a leaf. $\rd$-trees of height $\le e$ are called {\bf $(d,e)$-trees}.

Any map $\omega : \rd \rightarrow \fp$ defines a path in $T$ by answering the queries
by $\omega$. The label of the leaf on the path defined by $\omega$ is denoted $T(\omega)$. 

If $\omega$ is a linear map but not a homomorphism there must be a {\bf conflict pair}:
a pair of polynomials $g, g'\in \rd$, such that
$deg(g g') \le d$ and
$$
\omega(g) \cdot \omega(g') \neq \omega(g g')\ .
$$

The following notion is key to our approach.

\begin{definition}[\cite{Kra-finitary}] \label{pseudo}

For any $d, e \geq 0$ and $1 \geq  \gamma \geq 0$,
a $(d,e,\gamma)$-{\bf solution} of $\cal F$
is a non-empty finite set $\Omega$ of maps
$$
\omega\ :\ \rd \rightarrow \fp
$$
satisfying the following three conditions:
\begin{enumerate}
	
\item $\omega$ is $\fp$-linear map: 
$\omega(a)=a$ and $\omega(g) + \omega(h) = \omega(g + h)$, 
for all $a \in \fp$ and all $g,h \in  \rd$.
	
\item For any two polynomials $f \in {\cal F}$ and $g \in \rd$ it holds that
$\omega(f g) = 0$, assuming $deg(f g) \le d$.
	
\item 
For any $(d,e)$-tree $T$:
$$
\prob_{\omega \in \Omega}[T(\omega) \mbox{ is {\bf not} a conflict pair for $\omega$ }] \geq \gamma\ .
$$
\end{enumerate}
A {\bf pseudo-solution} is a collective name for  
$(d,e,\gamma)$-solutions.

\end{definition}

\begin{theorem}[{\cite[Thm.3.2]{Kra-finitary}}] \label{finitary}
	
	{\ }
	 
Assume that there exists an ENS-refutation of $\cal F$ with $S$ extension polynomials,
of degree $d$ and accuracy $h$ satisfying $e^{h/p}\geq 2S^2$.
Then there is no $(d, h + \log S, S^{-1})$-solution of $\fn$. 
\end{theorem}

The following theorem was pointed out 
and explained after \cite[Thm.5.2]{Kra-finitary} 
for families of polynomials systems $\fn$, $n \geq 1$,
containing all polynomials $x^2 - x$ for all $x \in Var(\fn)$
and satisfying that for some constant $c \geq 1$ it holds:
$$
|{\fn}|\le n^c\ ,\ \ |Var(\fn)|\le n^c\ \mbox{ and }\ 
deg(f)\le c\ \mbox{ for all } f\in \fn\ .
$$
However, it was not stated and proved there formally so we do it
now and we specialize again to system $\pphp_n$.

\begin{theorem} \label{reduction}
{\ }

For any constant $\ell \geq 2$ and any function $k = k(n)\geq 1$
the followings holds: if
there is an $((\log k)^{O(\ell)}, O(\log k), k^{-O(1)})$-solution for $\pphp_n$
then $\pphp_n$ does not admit 
$F_\ell(\mbox{MOD}_p)$-refutation with $k$ steps, for all $n >> 1$.

The constants implicit in the O-notation depend only on $F$ and $p$.
\end{theorem}

\prf

We follow the outline of the counter-positive
argument given at the end of \cite[Sec.5]{Kra-finitary}.
The passage from an $F_\ell(\mbox{MOD}_p)$-refutation of $\pphp_n$ to the non-existence
of a suitable pseudo-solution of $\pphp_n$ uses as an intermediary step proofs in ENS.

If we start with an $F_\ell(\mbox{MOD}_p)$-refutation of $\pphp_n$ with $k$ steps, 
Theorem \ref{bikprs}
gives us an ENS-refutation with:
$S = k^{O(1)}$, $\ell + O(1)$ levels and degree
$d \le (2 + \log k)(h+1)^{O(\ell)}$.

Choosing the accuracy $h$ minimal such that $e^{h/p}\geq 2S^2$, i.e. $h = O(\log S)$,
gives the bound to the degree
$$
d \le (\log k)^{O(\ell)}\ .
$$
Theorem \ref{finitary} then implies that there is no 
$((\log k)^{O(\ell)}, O(\log k), k^{-O(1)})$-solution of $\pphp_n$.

\qed

\section{Pseudo-solutions for the PHP} \label{php}

What we want is to construct 
for every $\ell \geq 2$ a $((\log k)^{O(\ell)}, O(\log k), k^{-O(1)})$-solution for
$\pphp_n$ with $k(n) = 2^{n^{-\delta}}$ for some $\delta > 0$
(but any $k(n) \geq n^{\omega(1)}$ would be interesting).
In fact, candidate pseudo-solutions were defined already in \cite[Sec.4]{Kra-finitary}
but we left it there as an open \cite[Problem 4.4]{Kra-finitary} to show that
they are indeed pseudo-solutions with the required properties.
In this section we shall
describe these candidate pseudo-solutions from \cite{Kra-finitary}.

\bigskip

To avoid a baroque notation put
$$
S(n,d)\ :=\ \phpd
$$
and define an $\fp$-vector subspace $V(n,d)$ of $S(n,d)$ to be
$$
V(n,d)\ :=\ Span[\{ g f \ |\ f\in \pphp_n \wedge deg(fg)\le d     \}]
$$
We will use the following theorem
that resulted from Razborov's proof \cite{Raz98} of degree $n/2$ lower bound for polynomial calculus PC refutations of $\pphp_n$.

\begin{theorem} [\cite{Raz98}] 
		{\ }
	
	For $2 \le d \le n/2$ it holds:
	\begin{enumerate}
	
\item $1 \notin V(n,d)$,

\item for any $g \in V(n,d)$ and $h \in S(n,d)$:
$$
deg(gh) \le d \ \rightarrow \ gh\in V(n,d)\ .
$$
\end{enumerate}
\end{theorem}
The second item implies that $V(n,d)$ is closed under degree $\le d$ PC derivations.
The first item yields the following statement.

\begin{corollary}\label{cor}
	{\ }
	
	For $2 \le d \le n/2$ there are maps $L : S(n,d) \rightarrow \fp$
	satisfying the first two conditions of Definition \ref{pseudo}:
	$L$ is $\fp$-linear, $L(1)=1$ and vanishes on $V(n,d)$.

\end{corollary}
Maps $L$ satisfying the conditions in the corollary 
 are usually called {\bf designs}
in proof complexity (cf. \cite{prf}) and we shall
denote their set by $\des(n,d)$.

\medskip

For a polynomial $g$ over $Var(\pphp_n)$ and a partial injective map 
$\rho : \subseteq [n+1] \rightarrow [n]$ define the {\bf restriction} $g^\rho$ of $g$ 
by $\rho$ the polynomial obtained by the following substitution
of $0/1$ values for some variables:

\[ x_{i j}^{\rho} =  \left\{ \begin{array}{ll}
	1       &  \mbox{if $i \in dom(\rho) \wedge \rho(i)=j$} \\
	0       &  \mbox{if $i \in dom(\rho) \wedge \rho(i)\neq j$} \\
	0       &  \mbox{if $j \in rng(\rho) \wedge \rho^{(-1)}(j)\neq i$} \\
	x_{i j}  & \mbox{otherwise}
\end{array}
\right. \]
Denote $D^{\rho} := [n+1] \setminus dom(\rho)$, 
$R^{\rho} := [n] \setminus rng(\rho)$ and
$n_{\rho} := |R^{\rho}| (= n - |\rho|)$. 
Note that
$$
x_{ij}^\rho = x_{ij}\ \Leftrightarrow\ (i,j) \in D^\rho \times R^\rho\ .
$$
For a set of polynomials $A \subseteq S(n,d)$ put
$A^\rho:=\{g^\rho\ |\ g \in A\}$.

\bigskip

Applying $\rho$ restricts the whole situation in the following sense. 
Substitution $\rho$ is a homomorphism of vector space $S(n,d)$ to 
$S(n,d)$
preserving multiplication
when defined. The restricted system 
$(\pphp_n)^\rho$ represents the negation of the PHP principle where
pigeons and holes are taken from $D^\rho$ and $R^\rho$, respectively, and
$S(n,d)^\rho$
is isomorphic to $S(n_\rho,d)$. Similarly, $V(n,d)^\rho$ is the set of polynomials
with a degree $\le d$ PC derivation from $(\pphp_n)^\rho$.

We shall denote by $\des(n,d)^\rho$ the set of degree $d$ designs on $S(n,d)^\rho$, to
keep a consistent notation when talking about a restriction.

\begin{definition}[{\cite[Def.4.2]{Kra-finitary}}]
	{\ }
	
For $2 \le d \le n/2$ let $\one$ be 
	the set of all maps $\omega\in \des(n,d)$ that are defined as follows:

\begin{enumerate}
		
\item Pick 
\begin{enumerate}

\item a restriction $\rho : \subseteq [n+1] \rightarrow [n]$ with $n_\rho = 2d$,
			
\item a map $L\in \des(n, d)^\rho$.
			
\end{enumerate}
		
\item For $g \in S(n,d)$ put:
$$
\omega(g)\ :=\ L(g^\rho)\ .
$$
\end{enumerate}
We shall write $\omega = (\rho_\omega, L_\omega)$ with $\rho_\omega$ 
and $L_\omega$ denoting the respective $\rho$ and $L$ defining $\omega$.
	
\end{definition}
Note that Corollary \ref{cor} implies that $\one \neq \emptyset$. 

\medskip

It appears rather difficult to analyze directly how hard it is to find 
by a tree a conflict pair
for maps from $\one$.  The following lemma 
reduces that task to a seemingly more transparent problem.

\begin{lemma}[{\cite[L.4.3]{Kra-finitary}}] \label{28.9.26b}
	{\ }
	
Let $T$ be a $(d,e)$-tree over $S(n,d)$. Then there is $(d,e')$-tree $T'$ with 
$e'\le e + O(d \log n)$ such that for any 
$\omega = (\rho, L) \in \one$ if $T(\omega)$ is a conflict pair
for $\omega$ then $T'(\omega)$ is a pair $(i,j) \in D^\rho \times R^\rho$.
\end{lemma}

The proof can be found in \cite{Kra-finitary} but its idea is simple:
tree $T'$ uses a conflict pair found by $T$ to find (by a binary search
argument) 
a conflict pair consisting of monomials and then another pair where one of the monomials
is a variable $x_{i j}$. Clearly then $\rho(x_{ij})\neq 0,1$, i.e.
$(i,j) \in D^\rho \times R^\rho$.

\section{Expressing the error of a tree} \label{key1}

Let $T'$ be a $(d,e')$-tree with labels in $[n+1]\times [n]$. 
Our aim in this section is to express the error
$T'$ must make on $\one$ in terms of restrictions only.
The phrase {\em $T'$ errs on $\omega = (\rho,L)$} means that 
$T'(\omega) \notin D^\rho\times R^\rho$

A path $P$
in $T'$ is a sequence of edges connecting the root with a leaf and $\labp$
denotes its label. 
For a path $P$ we define the set of polynomials $A(P)\subseteq S(n,d)$
consisting of all $g - a$ such that $g = ?$ is queried on $P$ and $P$ follows the edge 
corresponding to $g = a$.
Put $W(P) := \spa(V(n,d) \cup A(P))$.

When we apply a restriction $\rho$ to $T'$ a query $g = ?$ becomes $g^\rho = ?$,
and $A(P)$ and $W(P)$ become $A(P)^\rho$ and $W(P)^\rho$, respectively. Define:
$$
\dim^\rho(P) \ :=\ \dim(W(P)^\rho) - \dim(V(n,d)^\rho)\ .
$$
Sample $\omega =(\rho, L)$ will follow in $T'$ path $P$ and thus define $T'(\omega)=\labp$
iff $L$ vanishes on $A(P)^\rho$. Given $\rho$ such $L$ exists
iff $1 \notin \spa(W(P)^\rho)$ and in this case a random $L \in \des(n,d)^\rho$
will define $\omega := (\rho,L)$ following $P$ with probability 
$p^{- \dim^\rho(P)}$.

Let $\chi(P,\rho)$ be the characteristic function of the
set of pairs $P, \rho$ such that
\begin{equation}\label{chi}
\labp \notin D^\rho\times R^\rho\ \mbox{ and }\ 
1 \notin \spa(W(P)^\rho)\ .
\end{equation}
That is, $\chi$ is a $0-1$ function and it equals $1$ iff (\ref{chi}) holds.
The first condition in (\ref{chi}) means that for any $L$
$T'$ errs on $\omega = (\rho,L)$ 
following $P$ and the second one implies that for some $L$
$(\rho,L)$ indeed follows $P$.

Define a random variable with values in $[0,1]$ fo $\rho \in \des(n,d)$:
$$
\errho\ :=\ \sum_{P} \chi(P,\rho)\ p^{- \dim^\rho(P)}\ 
$$
where $P$ ranges over all paths in $T'$.

The first item in the following lemma summarizes the
discussion so far and the second item follows from the first one.

\begin{lemma}
	{\ }

For any $\rho\in \res(n,d)$ and $L$ chosen randomly from $\des(n,d)^\rho$:	
$$
\prob_L[T'\mbox{ errs on }\ (\rho,L)] = \errho\ 
$$	
and for $\omega$ chosen randomly from $\one$:	
$$
\prob_\omega[T'\mbox{ errs on }\ \omega   ] = {\mathbf E}_\rho[\errho]\ .
$$
\end{lemma}
The term ${\mathbf E}_\rho[\errho]$ is the expected value
$$
\sum_{\rho} r^{-1} \errho
$$
where $\rho$ ranges over $\res(d,n)$ and 
$r := |\res(d,n)|$,  
and using $\dim^\rho(P) \le e'$ we can lower bound this by
$$
\sum_\rho \sum_P r^{-1} p^{-e'} \ \chi(P,\rho)\ . 
$$
Assuming w.l.o.g. that all paths in $T'$ have the length $e'$ this
is just the expected value of $\chi(P,\rho)$ and as $\chi(P,\rho)$ 
has values $0, 1$ it is the probability
$$
\prob_{\rho,P}\ \chi(P,\rho)=1\ .
$$
Thus we have

\begin{lemma} \label{28.9.26a}
	{\ }
	
	Assume that every path in $T'$ has the length $e'$. Then
$$
\prob_\omega[T'\mbox{ errs on }\ \omega   ] \ \geq\ 
\prob_{\rho,P}\ [\chi(P,\rho)=1]
$$	
where $\omega \in \one$, $\rho\in \res(n,d)$ and path $P$ in $T'$ are 
chosen randomly and uniformly.
	
\end{lemma}

\section{Formulation of the reduction} \label{key2}

The following summary statement follows at once from Theorem \ref{reduction}
and Lemmas \ref{28.9.26b} and \ref{28.9.26a}.

\begin{theorem} \label{main}
	{\ }
	
	Let $\ell \geq 2$. 
Assume that for a $(d,e')$-tree $T'$ with
$$
d = (\log k)^{O(\ell)}\ \mbox{ and }\ 
e' = O(\log n) + O(d \log n) = d(\log k)\ 
$$
it holds that:
	\begin{equation} \label{todo}
		\prob_{\rho,P}\ [\chi(P,\rho)=1]\ \geq \ k^{-O(1)}
	\end{equation}
where $\omega \in \one$, $\rho\in \res(n,d)$ and path $P$ in $T'$ are 
chosen randomly and uniformly.
	
Then for all $n >> 1$ any $\fl$-refutation
of $\pphp_n$ requires at least $k(n)$ steps.  
\end{theorem}
(For the estimate of $e'$ by $d(\log k)$
we use that $k \geq n^3/2$ as 
any refutation of $\pphp_n$ must use all polynomials $Q_{i_1, i_2; j}$.)

\medskip

It appears possible that the hypothesis in the theorem holds for $k(n) = 2^{n^{\delta}}$,
for sufficiently small $\delta > 0$, even with the bound in (\ref{todo})
being $\Omega(1)$.

\section{A remark} \label{remarks}

The reader familiar with the Unstructured ENS proof system (UENS) from \cite{BIKPRS}
will notice that Theorem \cite[Thm.3.2]{Kra-finitary}
applies to UENS refutations of $\pphp_n$ too. Hence proving the hypothesis in 
Theorem \ref{main} with a super-polynomial $k(n)$
would imply that UENS does not simulate even $\mbox{TC}^0$-Frege systems as these do prove
$\php_n$ efficiently (cf.\cite{kniha}). This would disprove a casual remark 
at the end of \cite{BIKPRS} that UENS
simulates Extended Frege system EF. Note that it is unknown at present
whether EF simulates UENS.

\bigskip
\noindent

{\large {\bf Acknowledgments.}} A conversation with O.~Je\v zil helped me to
realize that I should write up the reduction and make it public.

\end{document}